# ESO-VLT BlueMUSE instrument - Conceptual Design for Phase A

Florence Laurent*[a], Johan Richard[a], Rémi Giroud[a], Davor Krajnović[b], Alexandre Jeanneau[a], Manuel Abreu[c], Angela Adamo[i], Théo Battrel[a], Didier Boudon[a], Zhemin Cai[e], Diane Chapuis-Kerouanton[d], Christopher Coote[e], Malak Galal[d], Joss Guy[f], Robert J. Harris[f], Andreas Kelz[b], Audrey Lanotte[g], Jon Lawrence[e], Rémy Le Breton[h], Helen McGregor[e], Gloria Mellinand[d], Jonathan Moller[e], Manuel Monteiro[j], Guiliano Parma[g], Arlette Pécontal[a], Peter M. Weilbacher[b]

[a]Universite Claude Bernard Lyon 1, CRAL UMR5574, ENS de Lyon, CNRS, Villeurbanne, F-69622, France
[b]Leibniz-Institut für Astrophysik Potsdam (AIP), Germany
[c]Instituto de Astrofísica e Ciências do Espaço, U. Lisboa, Portugal
[d]Ecole Polytechnique Federale de Lausanne (EPFL), Lausanne, Switzerland
[e]Australian Astronomical Optics, Macquarie Univ., Sydney, Australia
[f]Centre for Advanced Instrumentation, Department of Physics, Durham University, South Road, Durham, DH1 3LE, UK
[g]Dpt d'Astronomie, Univ. de Geneve, Versoix, Switzerland
[h]Universite Paris-Saclay, CEA, IRFU, Gif-sur-Yvette, France
[i]Stockholm University, Stockholm, Sweden
[j]Instituto de Astrofísica e Ciências do Espaço, U.Porto, Portugal

## ABSTRACT

BlueMUSE [1] is a blue-optimised, medium spectral resolution, panoramic integral field spectrograph under development for the ESO's Very Large Telescope (VLT). The project is now entering preliminary design phase. With an optimised transmission down to 350 nm, spectral resolution of R~3500 on average across the wavelength range, and a large FoV (1 arcmin²), BlueMUSE will open up a new range of galactic and extragalactic science cases facilitated by its specific capabilities.

BlueMUSE consists of several subsystems arranged along the light path. A calibration unit reproduces the VLT's optical conditions, while the fore optics reshape the telescope's focal image. The splitting and relay optics divide the field of view into 16 channels, each feeding an integral field unit that contains an image slicer, a spectrograph, and a detector vessel. The image slicer converts the 2D sub-field into a 1D pseudo-slit, which the spectrograph disperses into spectra recorded by a 4k × 4k CCD in each detector vessel. A vacuum and cryogenic system cools the detectors, and the data reduction software processes the raw data into data cubes which are subsequently processed by a data analysis software system. All subsystems are supported by the instrument main structure and enclosed in a thermal housing for stability. The whole instrument is managed by an integrated control system combining electronics and software.

This paper summarizes the baseline architecture, interfaces, and functional descriptions of the BlueMUSE instrument at the start of Design Phase. This architecture is derived from the top-level requirements and the experience acquired from MUSE. It presents the global concepts along with their preliminary performance estimates.

*florence.laurent@univ-lyon1.fr; phone +33 4 78 86 85 33; http://cral.osu-lyon.fr/

# 1. INTRODUCTION

BlueMUSE instrument [1] represents the next addition in wide-field integral field spectrograph at the ESO Very Large Telescope. BlueMUSE aims to reach near-UV/Blue wavelength and its concept is built on a balance between a proven design foundation provided by the MUSE instrument [2] and a number of key technological leaps that will help to open up a new range of galactic and extra galactic science.

This unique Integral Field Spectrograph combines a large field of view, a medium spectral resolution, an high throughput and observes down to 350nm. These characteristics combined, allow the instrument to cover a broad area of new science cases.

The instrument is developed by a consortium of 9 partners, organized around CRAL (Center for Research in Astrophysics of Lyon) as the PI-institute. An extensive feasibility study, called Phase A, has been performed from 2022 to 2024 that culminated to an official review in beginning 2025. Based on highly positive assessment from the ESO review board, an official construction agreement has been signed in 2026 between ESO and the BlueMUSE Consortium. BlueMUSE has now started its Preliminary Design Phase and will have its first light on the VLT in 2034.

This paper summarizes some key science cases, the project status, the current design concept and the high-level instrument performance.

# 2. SCIENCE CASE AND DERIVED INSTRUMENTAL REQUIREMENTS

## 2.1 BlueMUSE Science

BlueMUSE will open a unique view of the Universe by combining panoramic integral-field spectroscopy (IFS) with access to the blue optical domain. Such a combination will enable transformative studies across a broad range of astrophysical phenomena, from the physics of individual gas clouds and young massive stars to the assembly of galaxies and large-scale structures, covering the Solar Systems, the Milky Way and the Local Group, nearby galaxies and the distant universe. The well-matched combination of blue sensitivity, high resolution, and wide field-of-view positions BlueMUSE as an instrument capable of delivering transformative scientific discoveries, as well as a spectroscopic support complementing the data obtained by other facilities.

We present here three science programmes that will be enabled by BlueMUSE and have a potential to redefine our understanding of astrophysical phenomena.

- In the Milky Way and the Local Universe, BlueMUSE will open new avenues for studying stellar populations and their interaction with the surrounding medium. Spectroscopic surveys of young massive clusters and stellar nurseries will provide unprecedented insight into the formation, evolution, and feedback of massive stars, which are the primary drivers of the chemical and energetic evolution of galaxies. The instrument will enable detailed investigations of globular clusters, dwarf galaxies, and nearby nebulae, revealing their stellar content, chemical enrichment histories, and dynamical properties. At the same time, spatially resolved observations of ionised gas will trace the exchange of matter and energy between stars and the interstellar medium, offering a direct view of the processes that regulate star formation. BlueMUSE has also a potential to extend integral-field spectroscopy to studies of comets and interstellar visitors, linking galactic astrophysics with planetary science.

- For nearby galaxies, BlueMUSE will provide a comprehensive picture of how stars, gas, and dark matter shape galaxy evolution. Its sensitivity to key diagnostics of ionised gas in the blue wavelength range will allow measurements of physical conditions, chemical abundances, and gas kinematics, resolving the turbulent from gravitational motions. These observations will shed light on the circulation of gas within galaxies, the impact of stellar feedback, and the mechanisms that drive star formation and chemical mixing. BlueMUSE will be particularly powerful for the study of low-metallicity starbursts and low surface brightness galaxies, two classes of systems that probe extreme regimes of galaxy evolution and offer potential analogues to systems in the early

Universe. The instrument will also enable precise measurements of stellar kinematics and orbital structures, providing a fossil record of galaxy assembly, as well as the role of mergers and environment in shaping galaxies.

- At the Cosmic Noon, BlueMUSE will extend the legacy of deep IFS surveys by targeting the epoch when cosmic star formation reached its maximum ($2<z<3$). BlueMUSE will be able to reveal how galaxies acquire gas from their surroundings, how they regulate their growth through feedback, and how they interact with the circumgalactic medium. Deep observations will map diffuse gas around galaxies, trace the exchange of matter between galaxies and their environments, and uncover the physical processes that govern the emergence of large-scale structures. BlueMUSE observations of gravitationally lensed systems and extended Lyα nebulae will offer unique views of the earliest stages of galaxy and cluster formation, while also providing constraints on the sources responsible for cosmic re-ionisation.

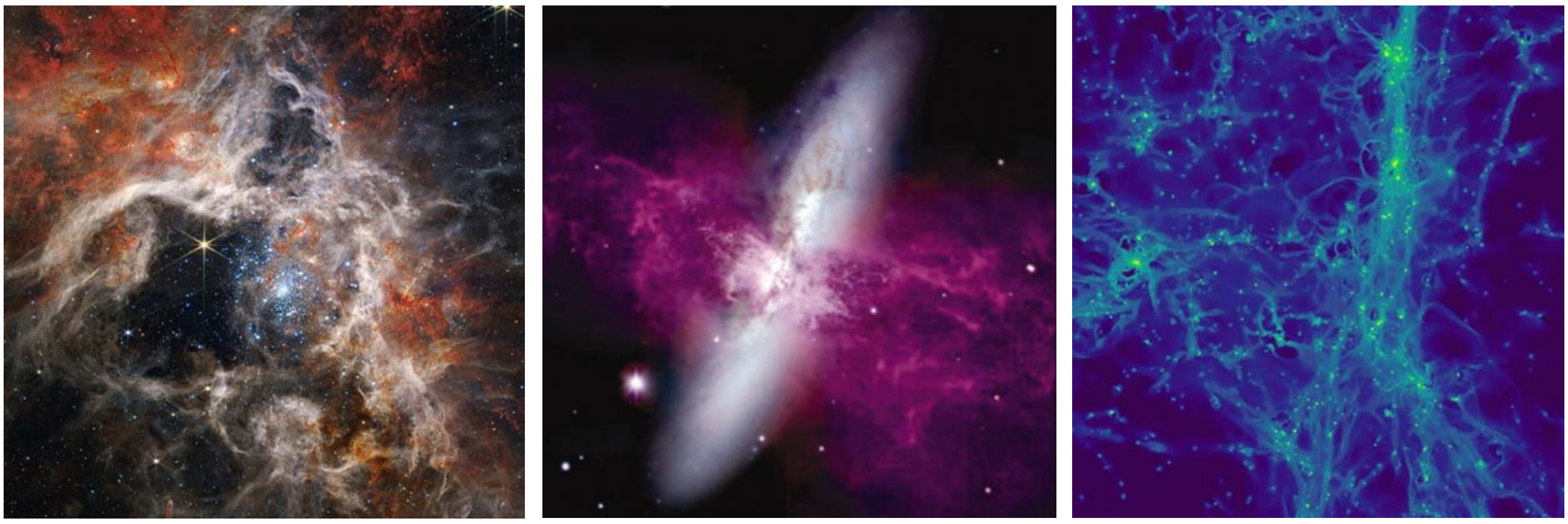

Figure 1. Three science programmes illustrated by three images. Left: the Tarantula nebula, a nursery of young massive stars. Middle: star-bursting nearby galaxy M82, featuring a super galactic wind of ionised gas outflowing from the central region powered by star formation. Right: a cosmological simulation of the distant universe showing the filamentary streams of hydrogen atoms linking galaxies in formation. Image credits from left to right: JWST/NIRCam - NASA (https://science.nasa.gov/asset/webb/tarantula-nebula-nircam-image/), HST/WIYN by NOIRlab – ESA (https://noirlab.edu/public/images/noao0404a/), and Project Sphinx – STScl (https://sphinx.univ-lyon1.fr/).

Taken together, these science possibilities establish BlueMUSE as a cornerstone facility for the coming decade. It will be both a "discovery" and a "follow-up" machine, working within the "spectroscopy-of-everything" concept, as it will obtain spatially resolved spectra for every source within its field-of-view. This will allow synergies with all major observatories operating at other wavelengths and spatial scales. Furthermore, by enabling studies of stellar populations and their properties, interstellar gas, circumgalactic environments, and large-scale structures such as groups, clusters and proton-cluster of galaxies, BlueMUSE has a potential to provide a coherent view of the evolution of galaxies and their constituents across cosmic time. In synergy with major observatories operating at other wavelengths and spatial scales, it will not only address long-standing questions in astrophysics, but also open discovery space for phenomena that remain beyond the reach of current facilities.

### 2.2 Key Top Level Characteristics

Hereafter is the list of some Top Level Requirements for the instrument. This unique combination of characteristics will allow BlueMUSE to contribute to all the Science Cases developed in the BlueMUSE White Paper (https://bluemuse.univ-lyon1.fr/wp-content/myimages/2025/06/BlueMUSE_Phase_A_White_Paper_submitted.pdf ).

Table 1. Key Top-Level Characteristics

| | |
|---|---|
| Wavelength Range | Nominal: 350nm - 580nm<br>Extended: 330nm-600nm |
| Spectral Resolution | Min >2600, Average >3500 |
| End to End Throughput | >15% (incl atmosphere, BlueMUSE and Telescope) |
| Field of View | 1 arcmin² |
| Image Quality | <0.42'' FWHM (w/o telescope) |
| Efficiency | >70% ratio between open shutter time and observation |
| Stability | Absolute difference of 0.1 pixel and illumination within 10% over one day |

It is to be noted that BlueMUSE will operated nominally in the wavelength range 350nm < λ < 580nm. However, we defined as "extended wavelength range" the additional wavelength acquired by the detectors but only partially covered by the field of view. This extended range covers approximatively 330nm < λ < 600 nm and a third of the field of view. All requirements related to the extended range are considered as "goals" at this point.

## 3. BLUEMUSE PROJECT

The consortium in charge of developing the instrument is composed of 9 institutes across 6 countries in Europe and Australia and is led by the Center for Research in Astrophysics of Lyon (CRAL).

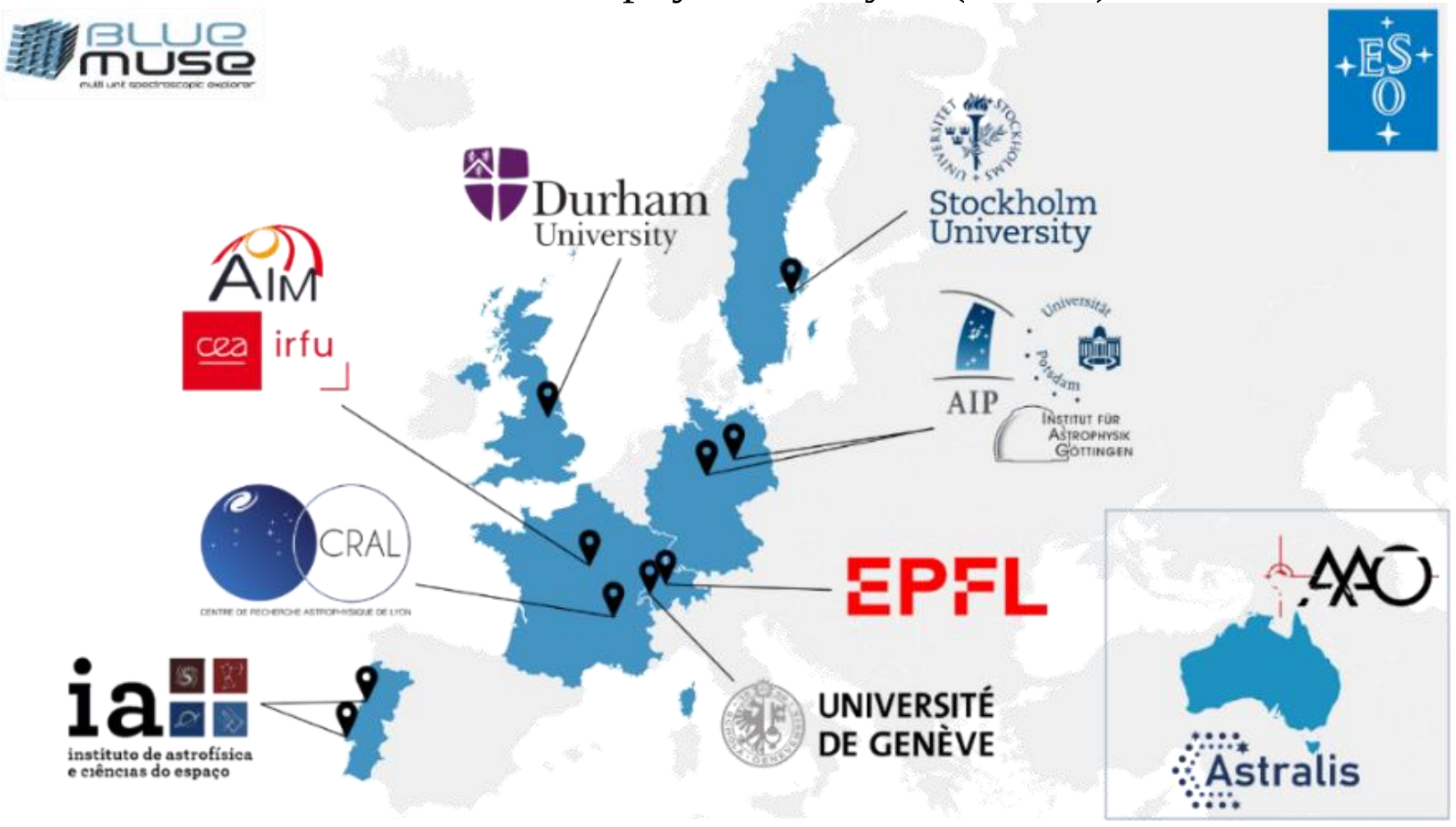

Figure 2. BlueMUSE Consortium Map

Each partner is responsible for one or several work packages, defined as below for the start of the design phase.

Table 2. Partner's responsibilities

| **Institute** | **Country** | **Responsibility** |
|---|---|---|
| CRAL - CNRS | France | PI-Ship, Image Slicer, System integration & test |
| AAO - MQ | Australia | Splitting & Relay Optics, Main Structure and Enclosure |
| AIM – IRFU - CEA | France | Detector system and Vacuum & Cryogenics |
| AIP | Germany | Calibration Unit and Science Software |
| Durham University | United Kingdom | Spectrograph and Integral Field Units |
| EPFL | Switzerland | Folding Mirrors Motorization |
| IA - CAUP - F.Ciências | Portugal | System Control Electronics and Software |
| Stockholm University | Sweden | Data Analysis Software |
| Université de Genève | Switzerland | Fore Optics |

The instrument is now officially in its construction phase that starts with a 3-years design phases (preliminary and final), followed by a 5-years MAIT phase in Europe and a final year at the Paranal Observatory with a first light planned in 2034.

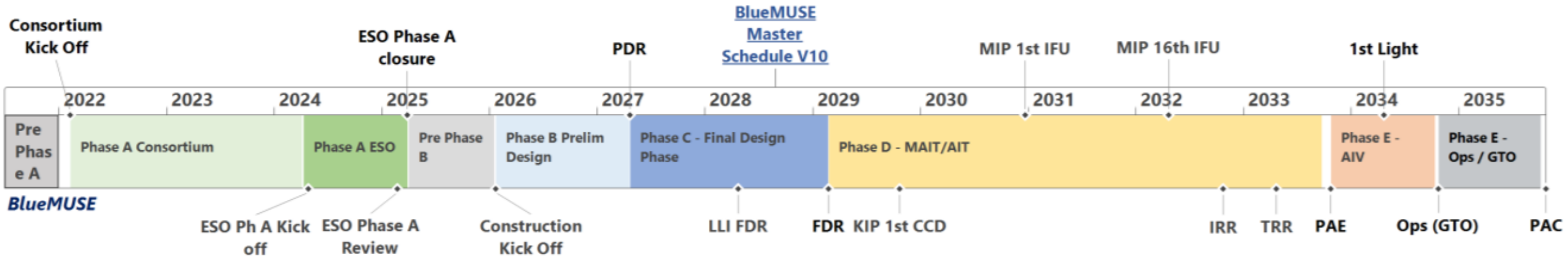

Figure 3. BlueMUSE Master Schedule

# 4. BLUEMUSE OVERALL DESIGN DESCRIPTION

## 4.1 System Description

The BlueMUSE concept is built around a modular and highly integrated optical system, where light follows a precisely optical path through multiple subsystems, each performing a specialized function to achieve the instrument's scientific goals.

At the start of the light path, the Calibration Unit (CU) plays a foundational role by illuminating the instrument with a calibration beam. This beam is carefully designed to mimic both the VLT's f-number and its illumination with different sources, ensuring that calibration processes accurately replicate real observational conditions. Next, the light enters the Fore Optics (FO) subsystem, which is responsible for reshaping the image delivered by the VLT. The FO corrects for field rotation, adjusts the magnification, and applies anamorphosis to transform the incoming light into a format suitable for the downstream optical path. The light then passes into the Splitting and Relay Optics (SRO) subsystem, where the field of view is divided into 16 separate channels. The SRO relays each channel's light toward the entrance of its corresponding Integral Field Unit (IFU), ensuring that the sub-fields are correctly directed and aligned. Within each channel, the Integral Field Unit (IFU) takes over, comprising three key components: the Image Slicer (ISS), the Spectrograph (SPS), and the Detector Vessel (DV). The Image Slicer performs a critical transformation by cutting the 2D sub-field of view into 48 individual slits and rearranging them into a single long pseudo-slit at the entrance of the spectrograph. The Spectrograph then disperses the light from each mini-slit, producing spectra that are imaged onto the detector The Detector Vessel houses the Instrument Detector System, which detects the dispersed light from the spectrograph. Each Detector Vessel contains a 4k x 4k CCD paired with its detector head and a cryocooler unit to maintain the detector at its optimal operating temperature. The detected light is then converted into digital signals and transmitted to the Data Reduction Software. This software removes the instrument's signature, corrects for environmental effects, and prepares the final data cube, which is a three-dimensional representation of the observed field, combining spatial and spectral information. To ensure the detectors operate under optimal conditions, the Vacuum and Cryogenics Support subsystem provides the necessary vacuum environment and cooling to the detector vessels. All these subsystems are structurally supported by the Instrument Main Structure, which serves as the backbone of BlueMUSE. This structure not only holds the subsystems in precise alignment but also interfaces with the Nasmyth platform of the VLT, ensuring mechanical stability and compatibility with the telescope. To further enhance stability, a passive thermal enclosure is installed around the instrument. This enclosure reduces external temperature variations during the night, minimizing thermal disturbances and improving the overall instrument stability. Given the complexity of BlueMUSE, with its multiple functions and devices, the instrument is also supported by a comprehensive control system. This system integrates both electronics and software control, enabling precise coordination of all subsystems, real-time monitoring, and automated adjustments to ensure optimal performance throughout observations.

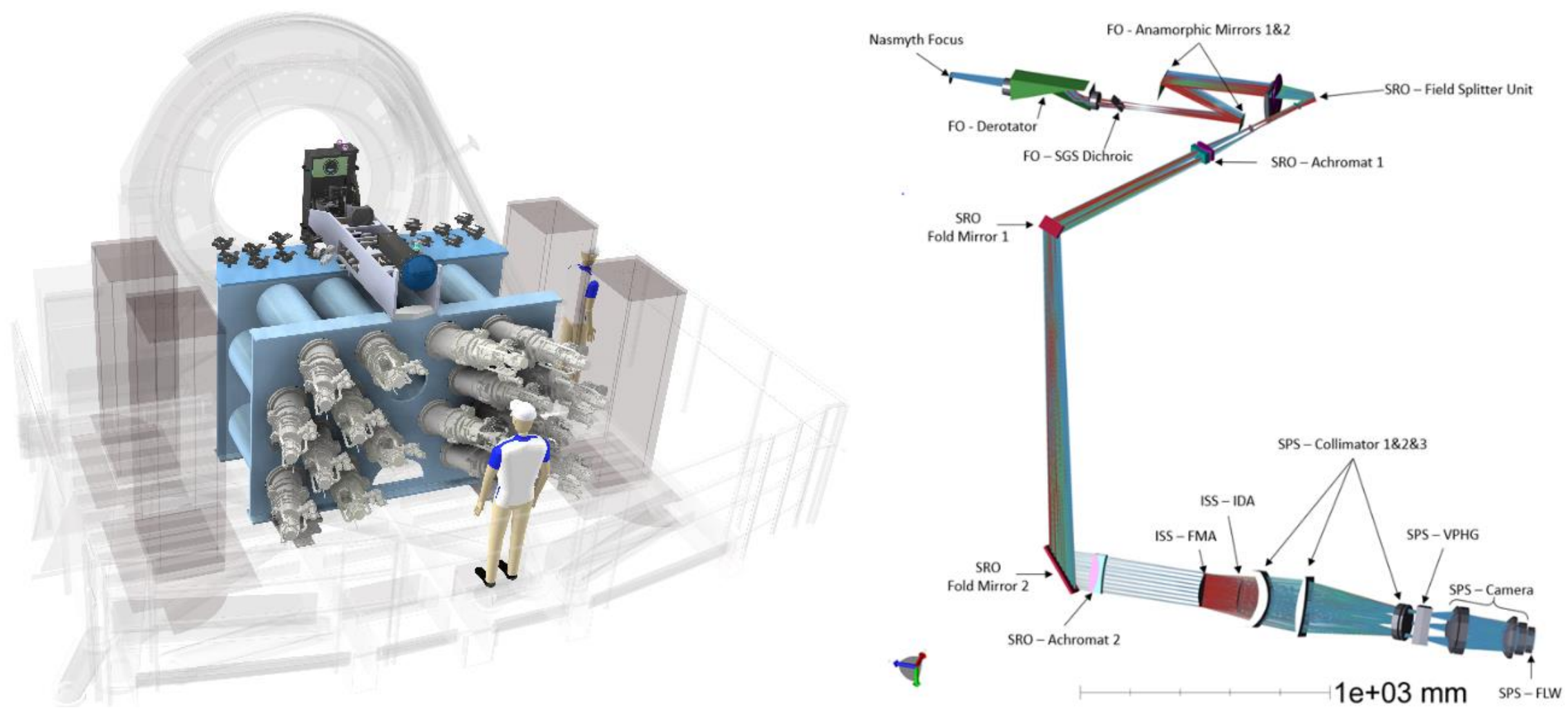

Figure 4. Left : CAD view of the overall BlueMUSE instrument on the Nasmyth platform. Right: End-to-end optical layout for channel 1.

## 4.2 Opto-mechanical Design

### 4.2.1 Calibration Unit

The Calibration Unit (CU) subsystem shall provide all means to fully calibrate the instrument either during daytime or during attached night-time calibrations. To achieve this, the CU delivers a uniform light beam that serves multiple calibration purposes. It provides illumination for flat-field calibration, ensuring that the instrument's response is consistent across its entire field of view. Additionally, the CU incorporates light sources specifically designed for spatial and spectral calibrations, enabling validations of the instrument's optical and spectroscopic performance. Furthermore, it includes field masks positioned in the instrument's focal plane, which are used for image quality assessments and geometrical calibrations. Moreover, the CU is its ability to mimic the Very Large Telescope (VLT) in terms of focal ratio and pupil shape. This ensures that the calibration process accurately replicates the observational conditions.

The design concept for the CU, proposed at the end of Phase A, presents several challenges, particularly concerning the light sources. BlueMUSE's requirements, especially its need for efficient and reliable light sources down to 330 nm, pose significant technical hurdles. To address these, the Leibniz Institute for Astrophysics Potsdam (AIP) is conducting extensive trade-off analyses to identify the most suitable solutions. For continuum light sources, the AIP is evaluating two primary options: a combination of Xenon and Halogen lamps versus a Laser-Driven Light Source (LDLS). For wavelength calibration, the team is comparing spectral arc lamps with a Fabry-Perot Etalon. Spectral arc lamps provide accurate, stable and unresolved emission lines, while a Fabry-Perot Etalon offers a tunable, more uniform and regular-spaced wavelength references.

Each option offers distinct advantages in terms of spectral coverage, stability, and efficiency, and the trade-off analysis aims to determine which best meets BlueMUSE's specifications.

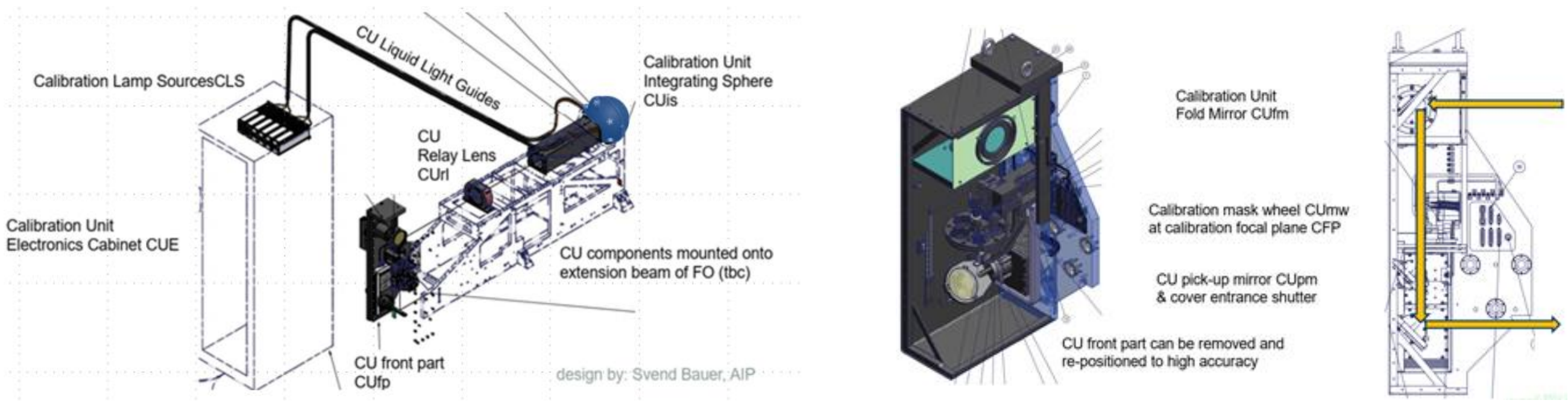

Figure 5. Left: Calibration Unit opto-mechanical components. Right: CAD view of the CU front part.

### 4.2.2 Fore-Optics

The Fore Optics (FO) subsystem serves as the interface between the telescope and the Splitting and Relay Optics (SRO) subsystem. Its primary role is to transform the rotating square field of view (FoV) delivered by the telescope into an anamorphosed, static FoV at the entrance of the SRO. In addition to reshaping the FoV, the FO subsystem provides dynamic information on any misalignment that may occur between the telescope and the derotated focal plane. This capability allows for real-time adjustments, ensuring that the optical alignment remains precise throughout observations. The FO also delivers a telecentric exit pupil, which is critical for maintaining uniform illumination. Furthermore, the FO enables synchronous, uniform, and adjustable exposure times in the image planes of all spectrographs simultaneously, ensuring that all channels capture data consistently.
The BlueMUSE Fore Optics is composed of several key components, each playing a specific role in achieving these functions:

- A derotator module, which includes two lens triplets and two prisms. This module compensates for the rotation of the telescope's field of view, ensuring that the FoV presented to the SRO remains static and properly oriented.
- A dichroic beam splitter, which selectively redirects the red visible portion of the incoming light toward a secondary guiding system (SGS). The SGS itself consists of a spherical triplet lens and a technical camera, which together provide precise guiding and alignment information to maintain the instrument's pointing accuracy.
- An exposure shutter, strategically positioned just after the dichroic beam splitter, at the location of the derotator's exit pupil. This shutter controls the exposure time for all spectrographs, enabling synchronized and uniform data acquisition across the entire system.
- An anamorphic system, comprising two cylindrical mirrors and a field toroidal lens. This system reshapes the FoV from its original square format into the anamorphosed configuration required by the SRO, ensuring optimal alignment and image quality.

Several improvements have been implemented in the BlueMUSE FO design to enhance performance and practicality. These include the use of a standard and relatively compact shutter, which simplifies the mechanical design while maintaining reliability. Additionally, the secondary guiding system now utilizes the full field of view in the red wavelength range, improving the accuracy and robustness of the guiding process. Finally, the overall design of the FO subsystem is more compact, reducing the physical footprint and integrating more efficiently with the rest of the instrument.

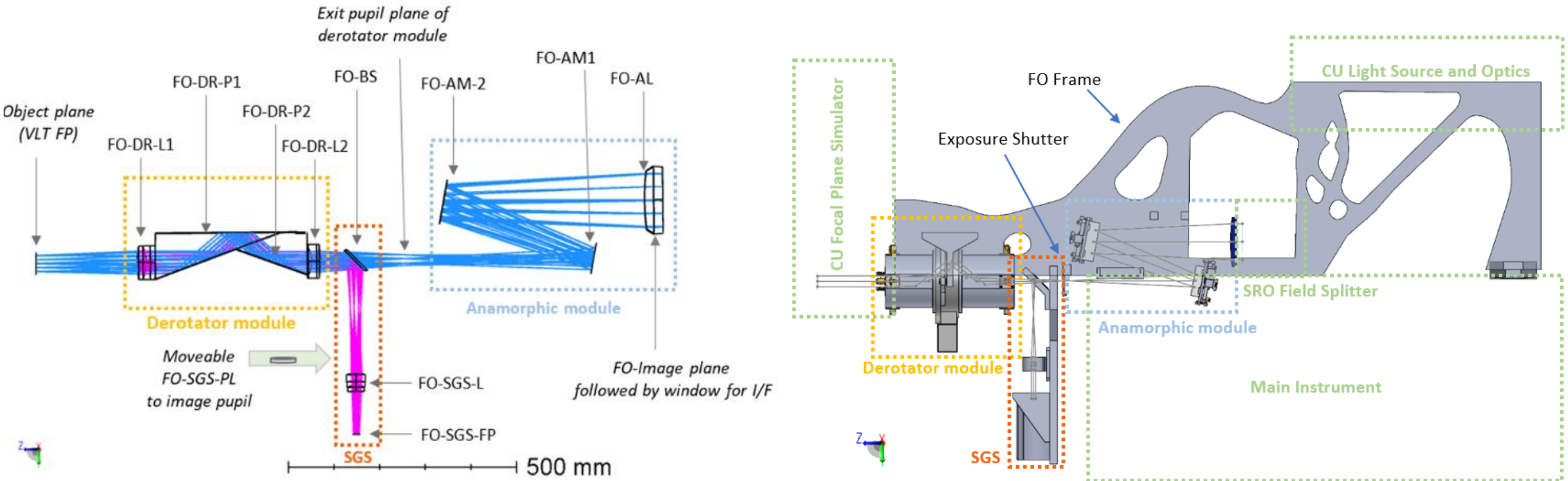

Figure 6. Left: Fore-Optics optical layout. Right: FO optomechanical design overview with main external subsystem interfaces (green)

### 4.2.3 Splitting and Relay Optics

The Splitting and Relay Optics (SRO) subsystem splits the field of view into sixteen sub-fields, as described in reference [3]. It relays these sub-fields from the Fore-Optics (FO) output to each Integral Field Unit (IFU) entrance. Simultaneously, it images the VLT pupil at infinity at the exit. The Field Separator Unit (FSU), a key part of the SRO, includes sixteen identical field lenses paired with their respective separator mirrors. The FSU is responsible for splitting the entrance FoV into 16 sub FoV, or channels, and direct each sub-FoV towards the relay units for further processing. Each of the sixteen relay units follows a precise optical path. This path begins with a first air-spaced achromat. Next, the light encounters two motorized 45° fold mirrors, also referred to as Folding Mirrors 1 and 2. These mirrors enable tip-tilt and focus re-alignment, as detailed in §4.2.4. Finally, the light passes through a bonded achromat. This configuration relays each sub-field of view toward its corresponding IFU entrance while applying the necessary corrections to achieve the required image size, quality, telecentricity, and chromatic aberration performance. The main engineering challenge lies in spatially arranging the 16 channels without any overlap. This must be accomplished while minimizing the overall volume to fit within the constraints of the Nasmyth platform. Additionally, the design must ensure thermal stability to prevent misalignment or performance degradation due to temperature fluctuations.

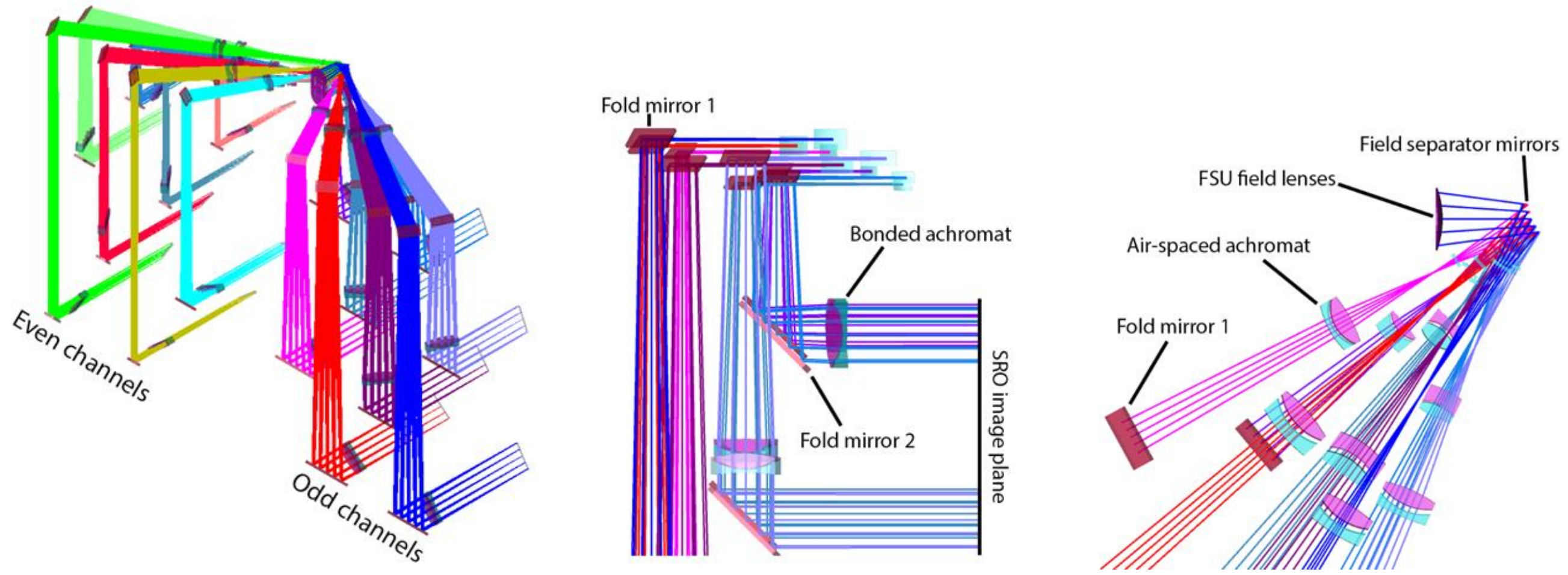

Figure 7. Left: Isometric full view of the BlueMUSE SRO optical layout. Middle: Upper paths of the SRO optical layout. Right: Lower paths of the SRO optical layout.

#### 4.2.4 Folding mirrors motorization

As part of the stability improvement activities during Phase A, the École Polytechnique Fédérale de Lausanne (EPFL) developed a motorized concept for the Folding Mirrors 1 (FM1) and Folding Mirrors 2 (FM2) mounts. This innovation is designed to compensate for seasonal instrument instability, ensuring long-term alignment and performance reliability.

The motorization system is intended for daytime calibration use only, activating only when a predefined displacement threshold is exceeded. This targeted approach minimizes unnecessary adjustments while maintaining optimal instrument stability.

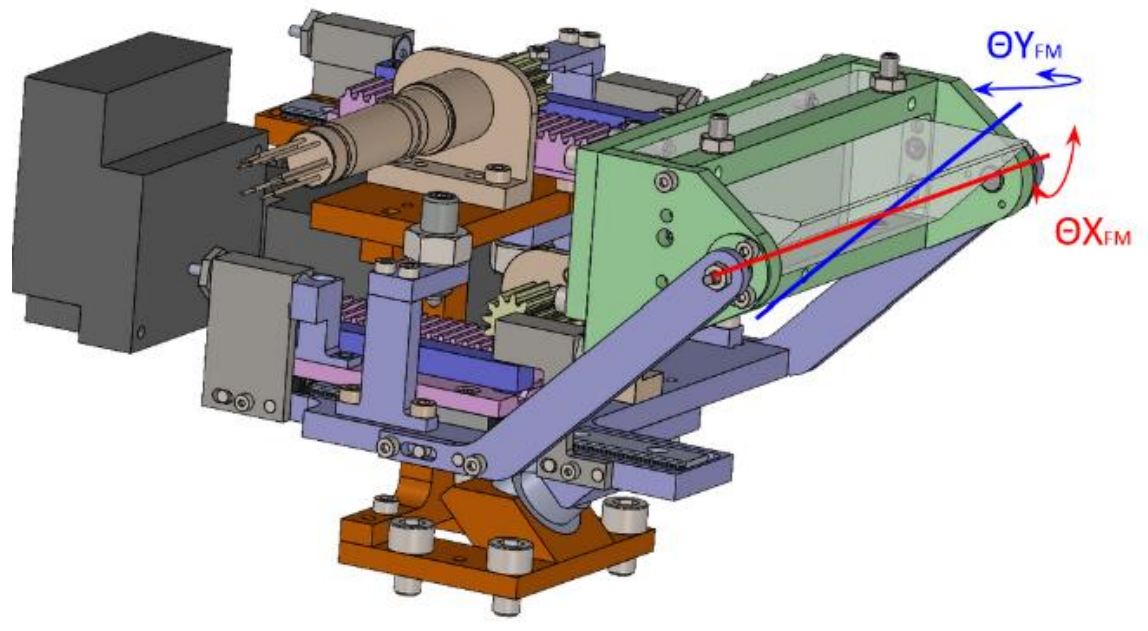

Figure 8. Overview of Folding Mirrors n°1 mechanical design

Both the FM1 and FM2 mount concepts are thoroughly documented in dedicated technical papers ([4] and [5], respectively). However, the introduction of motorization adds a significant number of motors (80 in total) to the system. This expansion necessitates a robust and proven control architecture, which is currently under development. The architecture will enable efficient and precise re-alignment during daytime operations, ensuring the system remains adaptable and responsive to environmental or mechanical shifts.

#### 4.2.5 Integral Field Unit

Each of the sixteen BlueMUSE Integral Field Units (IFU) includes an Image Slicer (ISS), a Spectrograph (SPS) and a Detector Vessel (DV).

- The IFU’s purpose is to obtain a spectrum of every point in a given contiguous sub-field of view.

- The Image Slicer transforms a rectangular field of view into a series of mini “slits” which are re-arranged along a pseudo-slit located at the entrance of the Spectrograph.
- The Spectrograph images the entrance slit at infinite distance and the entrance pupil onto a dispersive element. It then images all spectra onto the Detector.
- The Detector Vessel acquires and record the 16.8 megapixels of a raw exposures while maintaining the detector at its nominal operating cryogenic temperature.

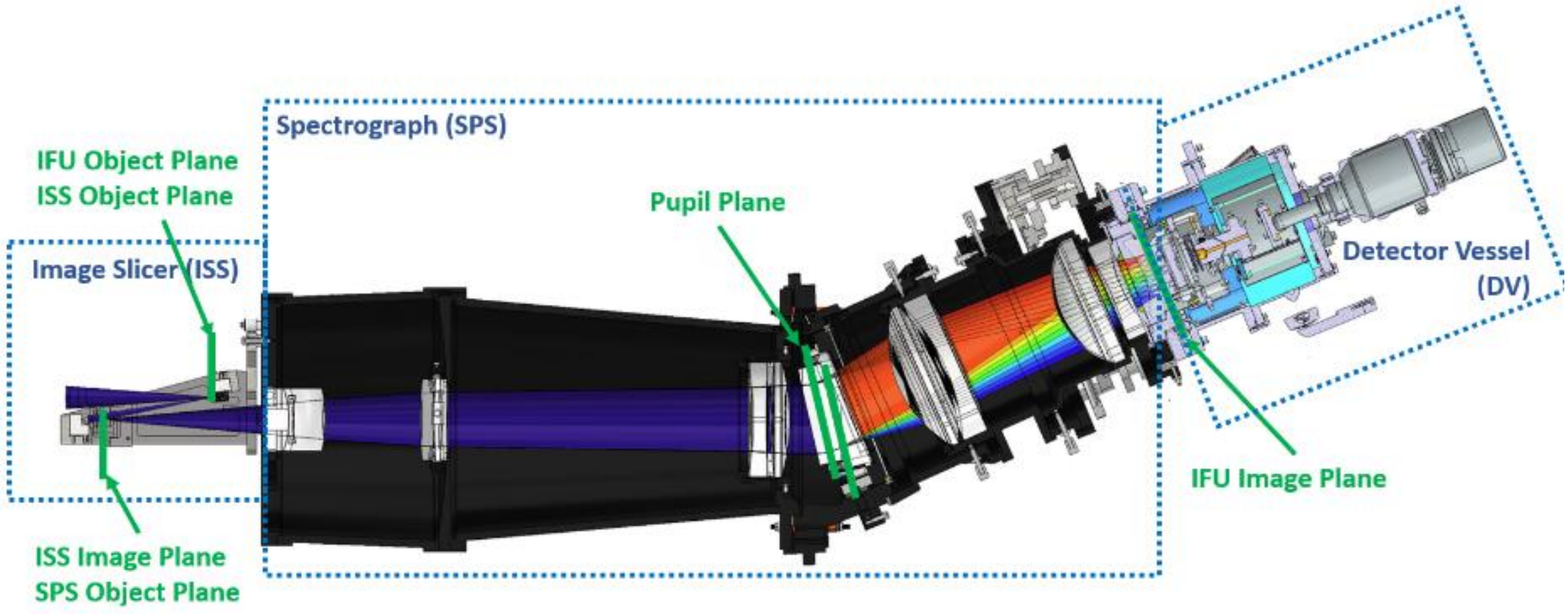

Figure 9. Overall views of the IFU

#### 4.2.5.1 Image slicer

The Image Slicer is a core component of BlueMUSE, responsible for dividing the entrance field of view (FoV) into 48 distinct optical paths and rearranging them into a 1D pseudo-slit for spectrographic analysis. It consists of three primary elements, each playing a critical role in ensuring optical precision and efficiency:

The first element is the Image Dissector Array (IDA), which comprises four identical stacks, each made up of 12 off-axis slices from a parent sphere. The IDA divides the entrance FoV into 48 separate optical paths while simultaneously imaging the telescope pupil in a brickwall pattern. This arrangement ensures that the light from each segment of the FoV is efficiently captured and directed toward the next optical stage.

The second element is the Focusing Mirror Array (FMA), which consists of four identical sub-arrays, each associated with an IDA stack. Every sub-array is composed of 12 tilted spherical mirrors, each corresponding to one of the slices in the IDA. The FMA performs three key functions: it images each slice into a "slitlet" at the image plane, images the telescope pupil at infinity, and ensures that the output chief rays of all slitlets are parallel. This alignment is crucial for maintaining the pupil and image quality across the entire field.

The third element is the Pupil and Slit Mask, positioned at an intermediate plane. This mask serves to reduce stray light, preventing unwanted reflections or scatter that could degrade the signal-to-noise ratio of the observations.

Given its complexity and the high precision required, the Image Slicer is one of the most critical and labor-intensive elements of BlueMUSE. To mitigate technical and cost risks, the team is leveraging the highly replicated and proven design from other projects, which has demonstrated its effectiveness in similar applications. However, BlueMUSE presents a unique challenge: with fewer Integral Field Units (IFUs) than MUSE, the Image Slicer must handle a faster beam along the spectral direction. To accommodate this, the FMA footprint has been slightly enlarged, though it remains within the spacing constraints imposed by the focusing mirrors.

As part of the Preliminary Design Phase, the team is also exploring cost-saving measures, including an alternative layout that reduces the number of slice stacks from four to three. This modification aims to simplify manufacturing and assembly while maintaining the optical performance required for BlueMUSE. The investigation will assess the feasibility, optical impact, and cost benefits of this change, ensuring that any redesign aligns with the instrument’s stringent performance criteria.

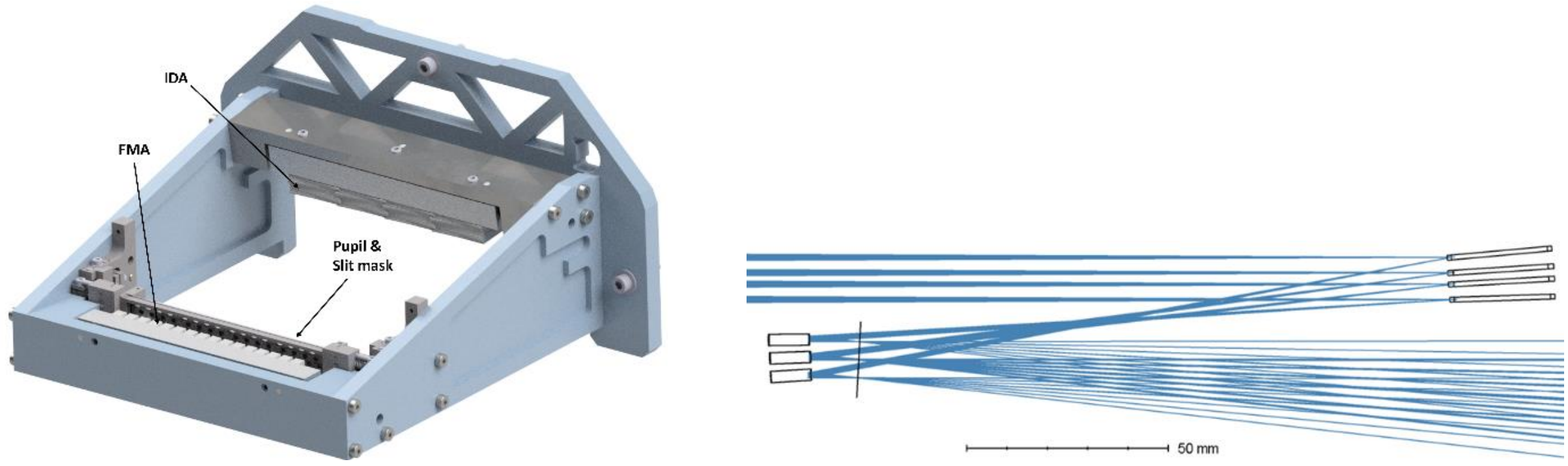

Figure 10. Left: ISS mechanical concept. Right: ISS optical layout. Only 4 slices are represented to improve readability

#### 4.2.5.2 Spectrograph

The BlueMUSE Spectrograph (SPS) is a precision-engineered optical system designed to efficiently disperse to efficiently disperse the light and to record the instrumental spectra, and it comprises three core modules: a Collimator, a Volume Phase Holographic Grating (VPHG), and a Camera [6].
The Collimator serves as the first optical stage, where it images the entrance pupil onto the VPHG while projecting the entrance pseudo-slit at infinity, effectively preparing the light for dispersion. Its design includes four lenses: two fused silica singlets and an air-spaced PBL35Y/CaF2 doublet, which together ensure high optical quality and minimal chromatic aberration. Notably, the front surface of the first lens and the back surface of the fourth lens are even aspheres, a feature that enhances the system's ability to correct for spherical aberrations and improve pupil quality across the field of view.
At the heart of the spectrograph lies the Volume Phase Holographic Grating (VPHG), a critical component responsible for dispersing the BlueMUSE wavelength range with high efficiency. The VPHG assembly is constructed with a hologram sandwiched between two 15 mm thick, plane-parallel fused silica plates, each measuring 150 x 125 mm. To further optimize performance, a spherical fused silica lens is bonded directly to the VPHG, ensuring mechanical stability and maximizing the throughput. Given its central role in determining the instrument's throughput and spectral resolution, the VPHG has been the subject of a dedicated study to evaluate different solutions and characterize their performance on a BlueMUSE testing bench [7].
The Camera is the final optical stage, where it focuses the dispersed spectra onto the detector with precision. It consists of five lenses: an air-spaced CaF2/PBL1Y doublet, two singlets (PBL1Y and fused silica), and a fused silica field lens that also serves as the detector vessel window. This configuration ensures high image quality and minimal distortion across the detector's surface.
BlueMUSE introduces several notable evolutions. These include physically enlarged optics to accommodate the instrument's expanded spectral sampling. Additionally, the Camera sub-assembly now features a motorized focus mechanism, allowing for fine adjustments to optimize image positioning to compensate for thermal effects between the operations. These enhancements collectively contribute to BlueMUSE's ability to deliver higher throughput, better spectral resolution, and greater operational flexibility.

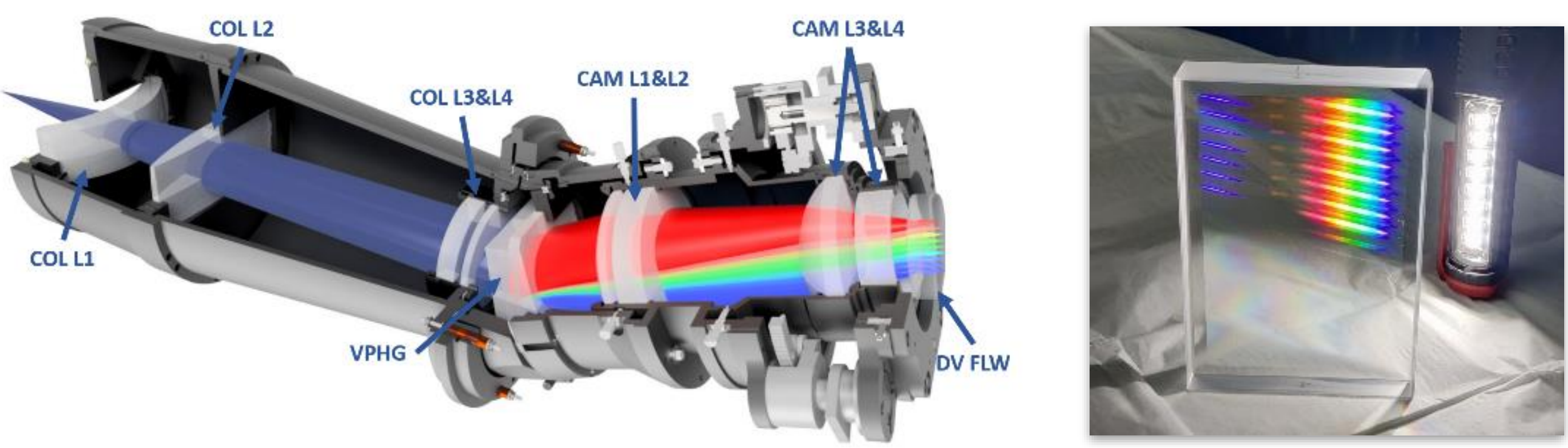

Figure 11. Left: I SPS Overview - Cross-section showing the optical elements and the beam. Right: VPHG prototype tested at CRAL 2 years ago.

#### 4.2.5.3 Detector Vessel

Each Detector Vessel (DV) in BlueMUSE is composed of a Cryocooler system and a Detector Head (DH), the latter serving as the protective housing for the 4k x 4k pixel, deep-depletion, backside-illuminated CCD with a 15 µm pixel size. The CCD's nominal operating temperature of 153 Kelvin is maintained by the Cryocooler, which replaces the traditional Continuous Flow Cryostat (CFC) system that relied on liquid nitrogen. This transition to a cryocooler-based system introduces key challenges, particularly in scaling the solution from one to sixteen units while managing mechanical vibrations and their potential impact on the instrument's structural stability.
To validate the design, two Detector Vessel Prototypes were constructed and tested for temperature control, vacuum integrity, and vibration levels. By the end of Phase A, the Cryotel GT was selected as the baseline cryocooler. However, the mechanical implications of deploying sixteen cryocoolers—especially concerning vibration propagation and its effect on the Instrument Main Structure—remain under active study. Additionally, discussions are ongoing regarding coating options for the CCDs to optimize performance.

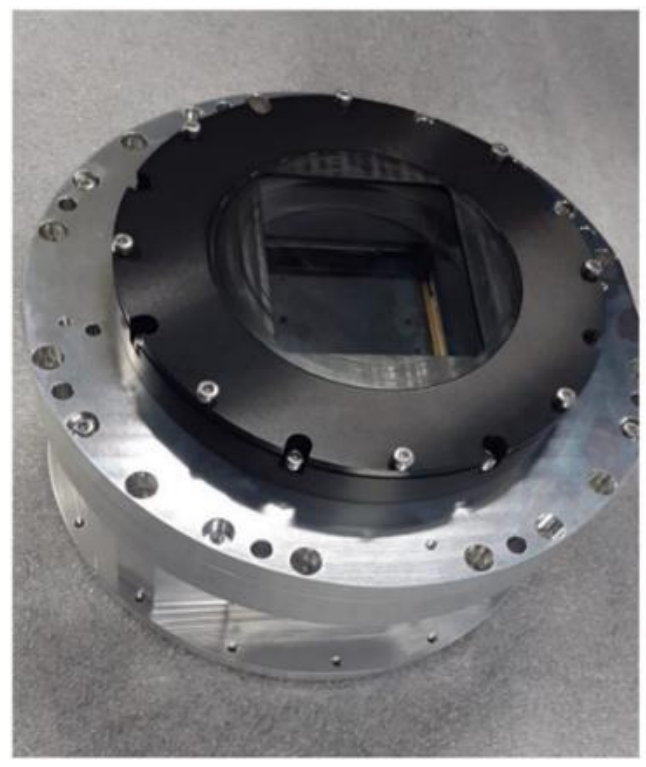
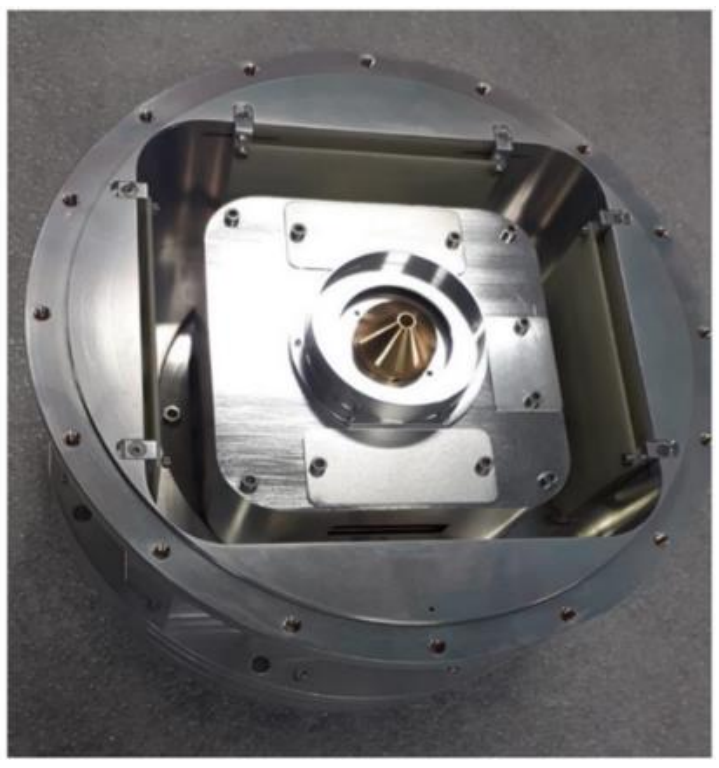
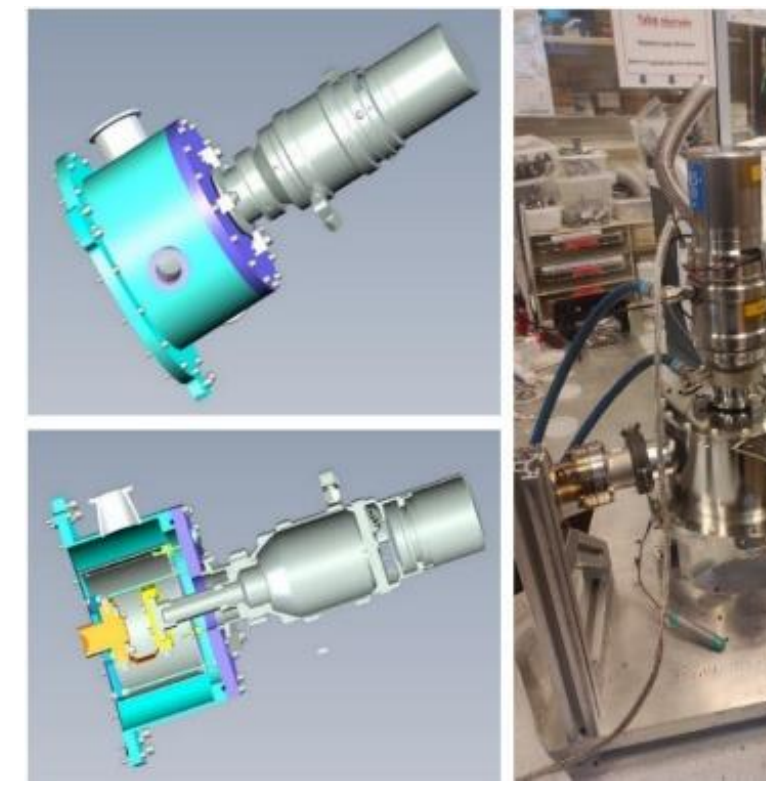
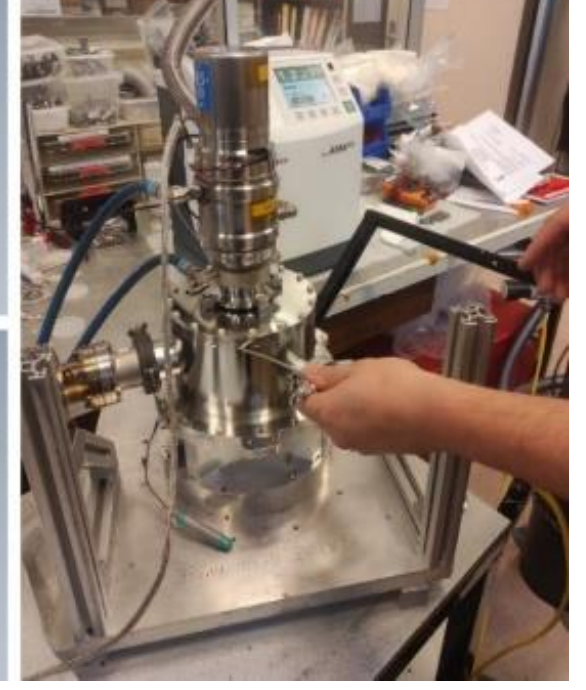

Figure 12. Left: Manufactured DH for the prototype DV: Front view and view of the inside with the conical interface (gold platted) and the thermal shield. Right: Cryotel GT- Design and fully assembled prototype, with DH at the bottom and Cryotel GT CH on top.

### 4.2.6 Structure and Enclosure

BlueMUSE's Instrument Main Structure (IMS) is a critical component designed for both structural integrity and thermal stability, and it is manufactured from a high-strength aluminum alloy. During Phase A, extensive studies and simulations [8] were conducted to optimize its performance, particularly to enhance temperature stability and minimize thermal distortions that could affect the instrument's optical precision.

The IMS architecture consists of front and back plates connected by cylindrical tubes, which are welded at both ends to ensure rigidity and alignment. Given the enlarged IFU (Integral Field Unit) in BlueMUSE, the diameters of these tubes were increased to accommodate the IFU's integration and maintenance demands. The entire IMS is securely supported by three kinematic mounts, which are bolted onto the Nasmyth platform rails, providing a stable interface with the telescope while allowing for precise alignment adjustments.

To further mitigate the effects of thermal variations—a critical factor for maintaining optical performance—the instrument will be fully or partially covered in a thermal enclosure. This enclosure is divided into two main sections: a front section and a rear section. The modular design allows for the partial removal of the rear section, such as during maintenance or upgrades, without disrupting the entire system. To streamline this process, the current concept includes one bulkhead connection panel on each side of the enclosure. These panels remain fixed to the Nasmyth platform even when the rear section is removed, ensuring that fluid lines (e.g., for cooling or vacuum systems) do not need to be disconnected during maintenance. This design minimizes downtime and reduces the risk of contamination or misalignment.

Until the Preliminary Design Review (PDR) phase, the team will continue to evaluate and refine different thermal management strategies. These studies will explore a range of concepts, including passive solutions (such as improved insulation or radiative shielding) and active control systems (e.g., thermal regulation via heaters or fluid loops). The goal is to determine the most effective approach to maintain thermal stability across the instrument's operational conditions, ensuring optimal performance for BlueMUSE's scientific objectives.

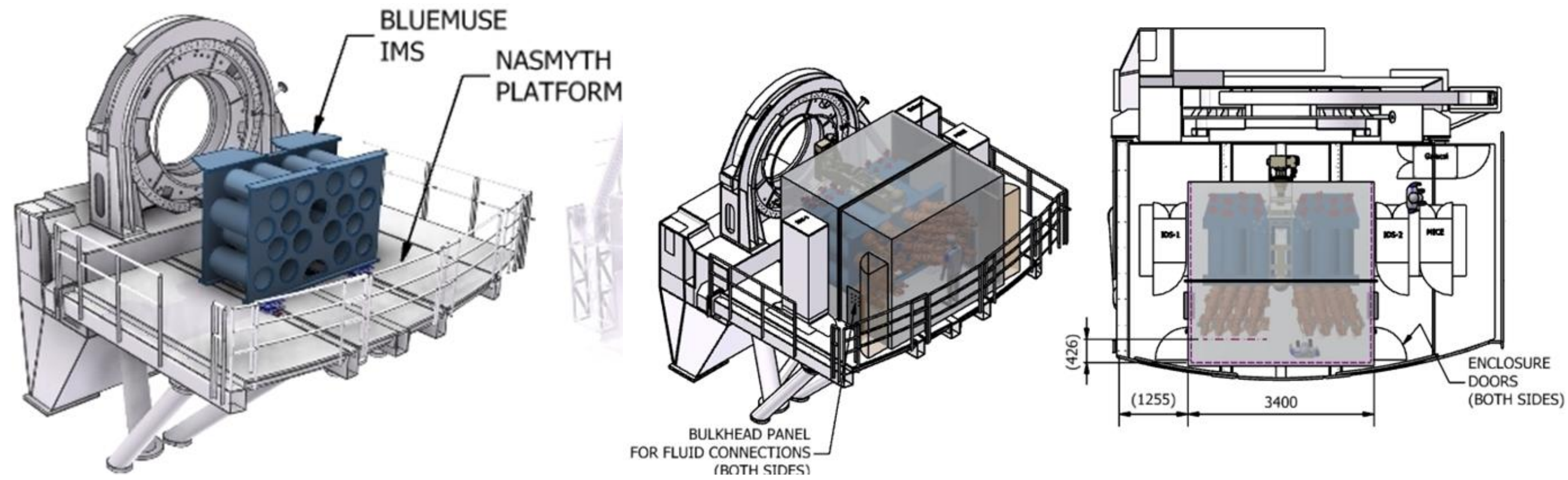

Figure 13. Left: Preliminary IMS isolated on Nasmyth Platform. Right: General dimensions of thermal enclosure (four cabinets not in their final position).

#### 4.2.7 Instrument Control Electronics and Software

The Instrument Control Electronics (ICE) and Instrument Control Software (ICS) subsystems are responsible for handling users' input and coordinating their executions.

To do so, BlueMUSE is expected to have two cabinets each equipped with a Beckhoff PLC. The control of the motorized folding mirrors, the possible detectors focus mechanism, the derotator drive, the various sensors data acquisition, the management of alarms and warnings, and general housekeeping tasks (cabinets temperature control, power supplies) will be split between those two PLCs.

The electronic architecture of the Folding Mirrors may span to the instrument structure with special hardware mounted on it. That architecture being quite heavy by handling eighty motors, it is described more thoroughly in the two papers about the Folding Mirrors [4][5] by EPFL.

In addition to the two previous PLCs, two additional cabinets will be used for the Detector Control System. They will include eight NGCII (New Generation Controller) connected to detector workstations that will control the scientific detectors. A third PLC for managing the Vacuum and Cryogenic System will also be placed in a separate cabinet. A Linux computer is also planned to be used to handle the technical camera using the ESO Frame Grabber Framework.

BlueMUSE ICS will rely on the ELT Instrument Control System Framework (IFW) meaning that a connection to an ELT-VLT gateway will be needed. The location of that gateway has yet to be determined.
The ICS relying on the ELT IFW, its architecture will follow the one made available and documented by ESO [9].

#### 4.2.8 Science Software

The BlueMUSE Science Software (BSS) subsystem includes three main components: the standard ESO observation preparation tools (including the Exposure Time Calculator), the Data Reduction System and the Data Analysis System.

The architecture of the Data Reduction System (DRS) builds upon the one available for the MUSE instrument and presented in detail in Weilbacher et al. [10] [11]. The data reduction workflow includes standard detector and spectrograph calibration at the channel level (bias, dark, flat, wavelength calibration) as well as global data reduction recipes (geometry, illumination, astrometry, sky subtraction). Data reduction is performed on individual BlueMUSE exposures and produces a calibrated datacube with a single interpolation step. In addition, specific improvements are currently developed in the context of BlueMUSE:

- We foresee the use of a new geometrical calibration algorithm, similar to the one presented by Piqueras et al. [12] (originally designed for the HARMONI instrument) and based on a simplified calibration sequence combining trace mask and multiphinhole mask exposures.

- The BlueMUSE optical design requires the pipeline to properly handle rectangular spaxels.
- A better estimate of the spectral and spatial variations of the Line Spread Function.
- The propagation of covariance information into the final data product.
- The possibility of combining BlueMUSE and MUSE datasets.

The large data volumes and rich information content delivered by panoramic IFS call for a modern and highly automated approach to data analysis. To address this need, we are developing a Data Analysis Software (DAS) framework that will support both BlueMUSE and MUSE datasets. It will provide a flexible yet largely automated workflow that users can adapt to their specific scientific objectives. Building on the output of the DRS, the DAS will generate science-ready data-cubes and high-level catalogues, including source identification, object classification, redshift determination, and measurements of emission-line properties, thereby enabling efficient exploitation of the data. In addition, selected products from the DAS may be fed back into the DRS to refine the reduction process, if necessary, and improve the quality of the final data-cubes. The BlueMUSE consortium is collaborating closely with ESO to define the software architecture and evaluate the integration of the DAS into the ESO data processing ecosystem.

# 5. INSTRUMENT PERFORMANCE

## 5.1 Image quality

We analyse the end-to-end model for channel 1, representing the worst-case scenario for image quality, and compute the spot RMS along the spatial and spectral axes. Note that the PSF along the spatial axis is influenced by all contributions from the sky plane to the detector, while the PSF along the spectral axis is shaped by contributions up to the image slicer. Assuming a Gaussian distribution for the effective contributions, we derive the FWHM from the spot size RMS. The resulting as-designed PSF FWHM along the spatial and spectral axes is presented in Figure 14. The spatial PSF remains within compliance (<0.42" FWHM), with comfortable margins on average across the common wavelength range.

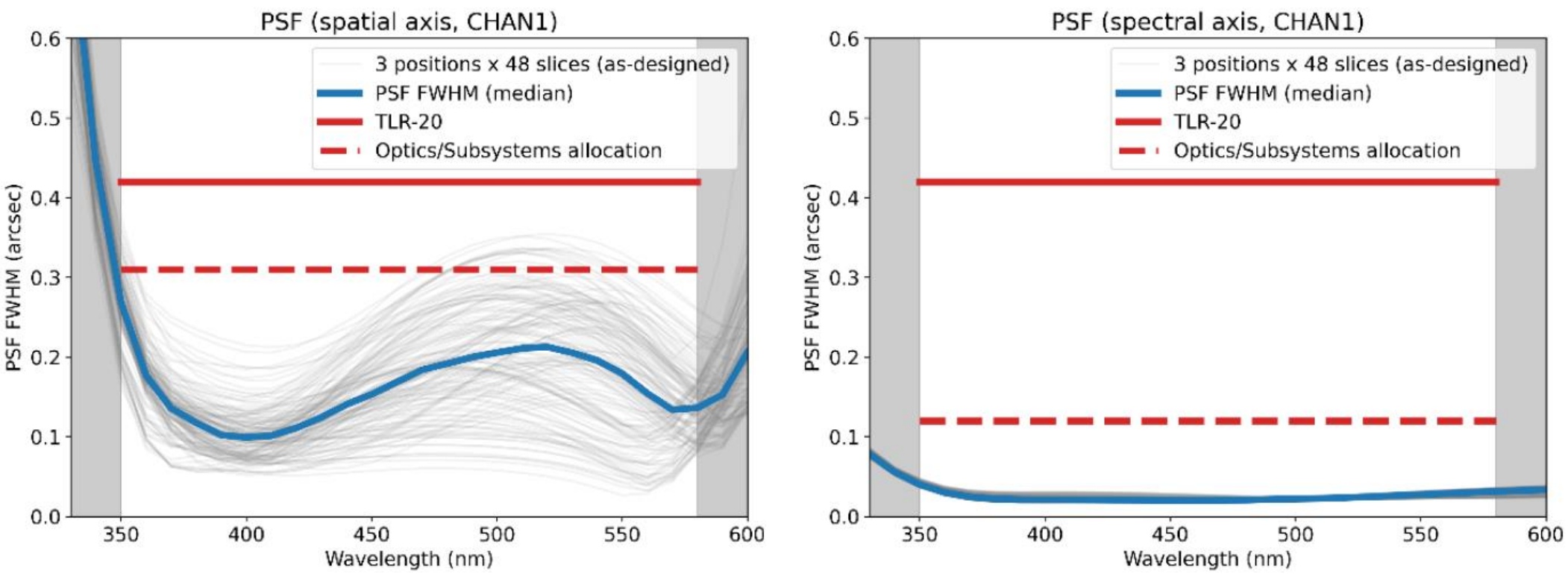

Figure 14. End-to-end PSF along the spatial (left) and spectral (right) axes

## 5.2 Spectral Resolution

The spectral resolution is determined by spectral sampling (LSF FWHM) and spectrograph dispersion (dλ/dpx), both derived from the IFU model for three field points per slice. The LSF FWHM is modeled as the convolution of a top-hat slit function with a Gaussian matching the measured spot RMS size. Analysis of MUSE's as-designed vs. as-built performance shows minimal LSF FWHM degradation, requiring fewer margins compared to the PSF. The Primary driver is the variation in projected slit height causes the LSF FWHM to exceed the budget at red wavelengths, prompting adjustments (e.g., spectrograph magnification) to introduce a margin (~2.00px instead of ~2.12px). The Spectral resolution is 2524 at 350 nm (below the requirement target of R > 2600) and averages 3507 (meeting requirement of R > 3500) - Figure 15. Non-compliance at minimum resolution is due to slight dispersion variation toward the blue, to be addressed in the next design phase.

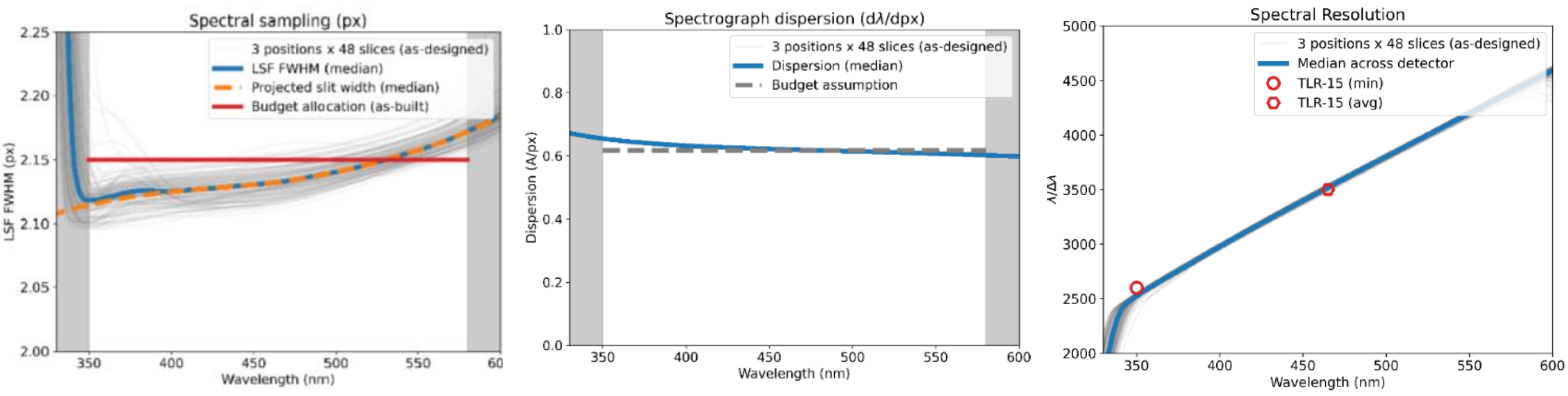

Figure 15. As-designed spectral sampling (left), spectrograph dispersion (middle) and as-designed spectral resolution (right)

## 5.3 End-to-end throughput

The transmission is computed along a single unpolarised ray at center field and edge of the pupil using the Zemax CODA operand. The default Anti-Reflective (AR) and High-Reflectivity (HR) dielectric coatings assume T or R = 99.3%, consistent with early LMA optimisations for $Ta_2O_5/SiO_2$ AR coatings in the 0–20° range, with margin for tolerances and dust losses. The cement transmission (e.g., in the derotator or field splitter) is not yet considered. In the Fore-optics, the dichroic is treated as an AR coating for now. In the Splitting and Relay Optics, the transmission is taken as the median of all SRO paths at each wavelength, with room for optimisation as the current design is likely over constrained for color correction and under constrained for transmission. The Image Slicer is based on Silflex UV measurements (optimized for KCWI). Slice shadowing losses (1–2% in MUSE) are not yet included. Focusing mirrors assume an HR coating with R = 99.3%. The VPHG diffraction efficiency is the current best estimate following the latest photopolymer-based VPHG optimisation, which allows trading red transmission for improved performance at 350 nm with minimal overall impact. The Detector Quantum efficiency curves are based on e2v (Astro Multi-15 and STA ηA, with broadband coatings. We model sky transmission at unit airmass with 2.5 mm precipitable water vapor using ESO SkyCalc. Telescope reflectance is based on fresh VLT mirror coatings ("Performance of the VLT Mirror Coating Unit", ESO Messenger, 97). We use this optimistic assumption to remain consistent with requirements.
The resulting system and end-to-end transmission curves are shown in the Figure 16, Left, along with current best estimates (CBE) and budget allocations for each contributor. The current best estimates only marginally meet the requirement, both at 350 nm and on average. Custom coatings, including graded coatings like MUSE's, could enhance transmission, especially at 350 nm.

## 5.4 Mass and Volume

The platform's maximum allowable mass is 8,000 kg. The current best estimate for BlueMUSE's mass is approximately 6,600 kg, providing a margin of over 20% compared to the specification. Given the varying design maturity levels across subsystems, we assigned a specific contingency percentage to each. When combined, the total applied contingencies amount to 19%, which is a reasonable assumption for a Phase A design.
The instrument's layout, featuring multiple Integral Field Units and a thermal enclosure, makes its footprint on the platform a critical consideration (Figure 16, Right). Currently, the front of the instrument encroaches into the restricted area, similar to MUSE. While adopting MUSE's solution of removing the front part of the Calibration Unit could work for BlueMUSE, a minor clash (~30 mm) remains at the top of the instrument. If ESO deems this critical, increasing the back focal distance—while preserving space for cryocooler maintenance—could be a solution.
Additionally, even after optimizing the cabinet layout and minimizing gaps, the design still clashes with a restricted access zone on the platform. A more detailed layout study during the PDR phase, in collaboration with Paranal, will determine if a platform extension is necessary. At this stage, no major blocking issues are anticipated, but a thorough footprint analysis at PDR will confirm any potential platform impact.

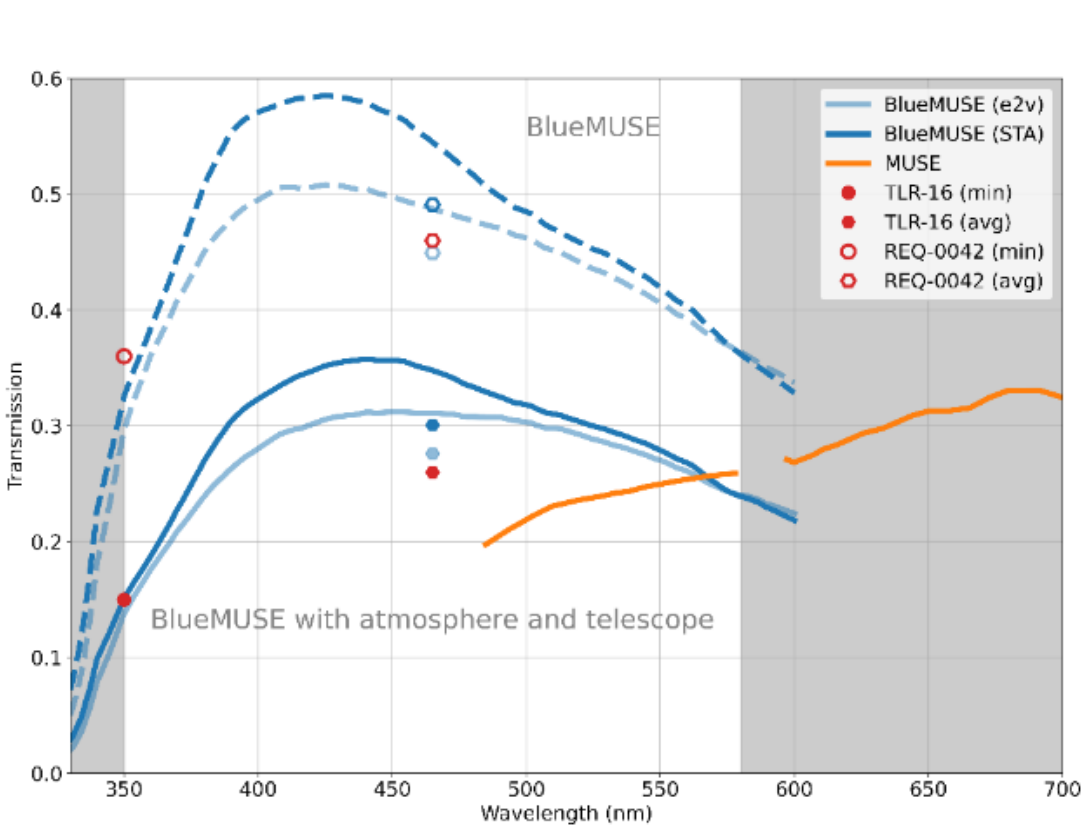

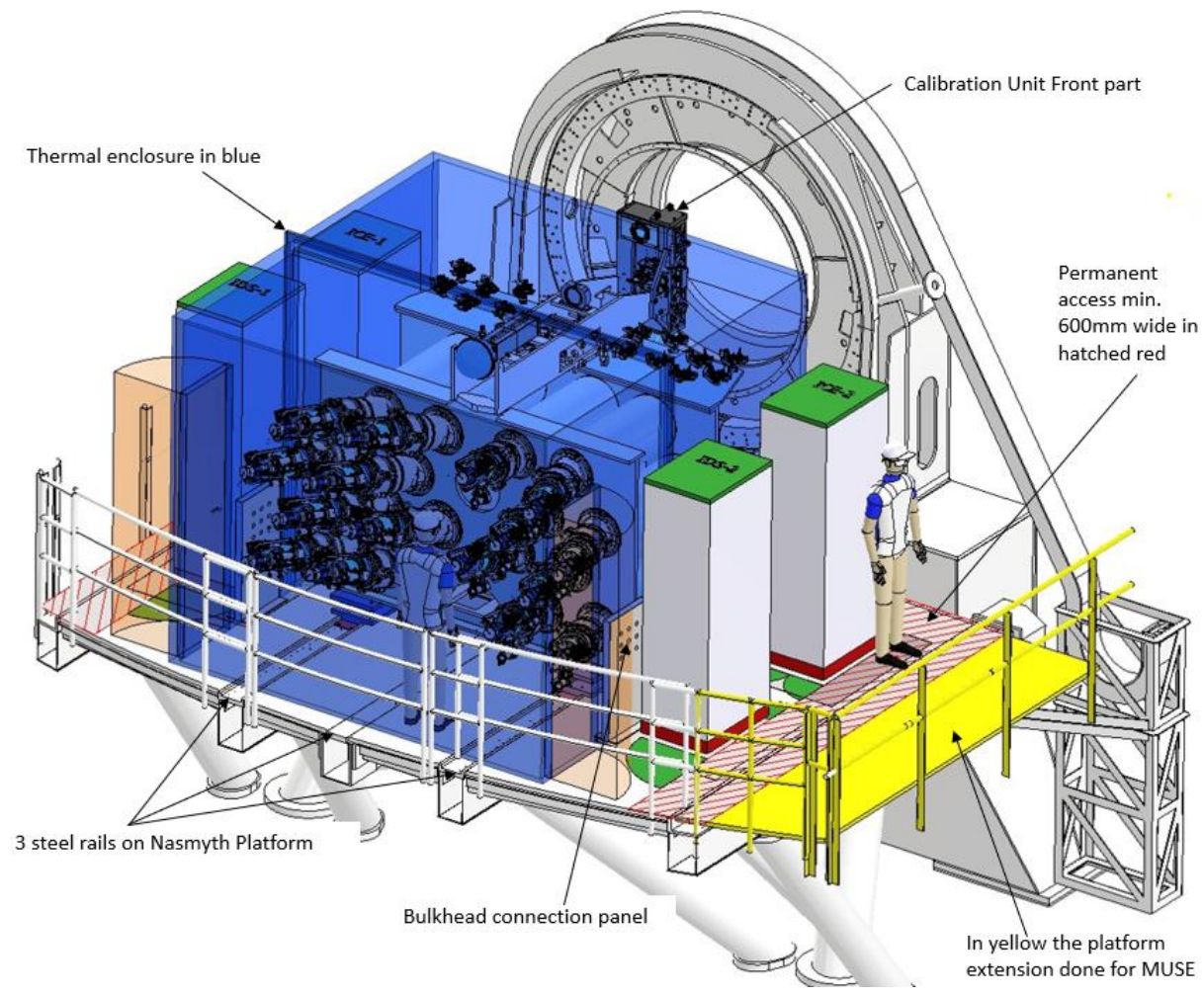

Figure 16. Left: BlueMUSE system and end-to-end transmission. The system transmission average over the 350-580 nm range is represented at 465 nm by empty hexagons, while the end-to-end transmission average is represented by filled hexagons. Right: BlueMUSE overall footprint on Nasmyth patform

### 5.5 Stability

A critical top-level requirement for the BlueMUSE instrument is stability: “in operational conditions, the BlueMUSE instrument shall be stable in spatial and spectral position to allow for the calibration of sky position and wavelength within an absolute difference of 0.1 pixel (including the effect of derotator wobble), illumination within 10%, based on calibrations taken during the day or twilight over the whole field of view and for the common wavelength range over one day.”

The design accounts for temperature variations of 1°C/hour, 5°C overnight, ~12°C monthly, and 0–20°C annually. The stability budget breaks down into four parts. The pipeline accuracy is based on QC parameters, with new algorithms under development for precise calibrations. The moving parts includes the Secondary Guiding System residues coming from the derotator wobble. The subsystem stability, the most critical part, addressing temperature and deformation effects on field position and wavelength across six subsystems That requires a balance between design choices and manufacturability. The forth parts is the margins, following Phase A standards. The stability is categorized into short-term (night/day) and long-term (monthly). This work is ongoing, and a baseline will be proposed for the PDR.

## 6. CONCLUSION AND FUTURE DEVELOPMENTS

BlueMUSE represents the next step of panoramic Integral Field Spectrograph on ESO’s Very Large Telescope. With a combination of a large field of view (1arcmin²), a medium spectral resolution (R=3500 on average), a strong efficiency (one observation mode) and an optimized transmission down to 350nm, the instrument will be a quite unique “discovery” and “follow-up” machine for various science cases ranging from investigation of the solar system and the Milky Way, to the Local Group and nearby galaxies up until the investigation of the distant Universe.

During Phase A, the consortium, in collaboration with ESO, has proven that BlueMUSE will be a key contributor for Science on the VLT in the 2030s and beyond. The instrument has well defined science cases, clear performance requirements, a challenging but proven design concept and a secure construction plan.

For all the technical challenges, a combination of prototyping testing and analysis during Phase A allowed the project to reduce the risk and improve the instrument performance.

The next step is the Preliminary Design Review planned for 2027, where the project shall demonstrate the design concept maturation and a clear strategy for next phases.

## ACKNOWLEDGMENTS

The BlueMUSE consortium is grateful for the following support:

- CNRS/INSU for their continued support through CSAA and National Programs. We also acknowledge support from Université Claude Bernard Lyon 1 / Observatoire de Lyon.
- The German contributions are funded by AIP, IAG, UP and by the BMFTR ErUM framework programme.
- The UK contribution is supported by funding from Durham University and from the Science and Technology Facilities Council under grant ST/Y000021/1X.
- The Australian contribution is supported by the Australian Government's National Collaborative Research Infrastructure Strategy (NCRIS).
- The Swiss contribution has been supported by the Funding LArge international REsearch projects (FLARE) program, through the grants 20FL21_201777, 20FL20_216690 and 20FL20_23694.
- The Portuguese contribution was supported by Fundação para a Ciência e a Tecnologia (FCT) through national funds under the research grants UID/04434/2025 (DOI: 10.54499/UID/04434/2025).